\documentclass[aps,prb,twocolumn,showpacs]{revtex4}
\usepackage{amsmath,amssymb,graphicx}
\begin{document}

\title{Ultrafast resonant optical scattering from single gold
nanorods: \\ Large nonlinearities and plasmon saturation}

\author{Matthew Pelton}  \email{pelton@uchicago.edu}
\author{Mingzhao Liu}
\author{Sungnam Park}
\altaffiliation{Current address: Department of Chemistry, Stanford
University, Stanford, CA  94305}
\author{Norbert F. Scherer}
\author{Philippe Guyot-Sionnest}

\affiliation{Department of Physics, Department of Chemistry, and
James Franck Institute, University of Chicago, 5640 S. Ellis Ave.,
Chicago, IL 60637}

\date{\today}

\begin{abstract}
We measure nonlinear optical scattering from individual Au nanorods
excited by ultrafast laser pulses on resonance with their
longitudinal plasmon mode. Isolating single rods removes
inhomogeneous broadening and allows the measurement of a large
nonlinearity, much greater than that of nanorod ensembles.
Surprisingly, the ultrafast nonlinearity can be attributed entirely
to heating of conduction electrons and does not exhibit any response
associated with coherent plasmon oscillation. This indicates a
previously unobserved damping of strongly driven plasmons.
\end{abstract}


\maketitle

\section{Introduction}

Conventional photonic devices are restricted by the diffraction
limit to be larger than half the optical wavelength, limiting the
possibilities for miniaturization.\cite{Saleh91} One way to overcome
this limit is to couple light to material excitations. Of particular
interest are collective electron oscillations in metal
nanostructures, termed surface plasmons.\cite{Barnes03,Maier05}
Plasmons allow for nanoscale delocalization, transport of
electromagnetic energy, and large local field enhancements.
Preliminary steps have been made, for example, towards using surface
plasmons in nanoparticles to construct sub-wavelength
waveguides.\cite{Quinten98,Maier03} Actively controlling light
propagation in such structures will require an understanding of the
ultrafast nonlinear response of the individual elements. Such an
understanding is also crucial for the treatment of effects such as
surface-enhanced Raman scattering,\cite{Nie97,Kneipp97} since it may
limit the magnitude of local electromagnetic fields that can be
achieved in real structures.

Previous measurements of metal-nanoparticle nonlinearities have
generally involved excitation at frequencies away from the plasmon
resonance, and thus probe incoherent effects related to the heating
of
electrons.\cite{Feldstein97,Hodak00,Voisin01,Link03,Arbouet03,Link00,Hu03}
Exciting and probing nanoparticles on resonance with their plasmon
frequencies can reveal nonlinearities associated with the coherent
oscillation of the plasmons themselves. Unfortunately, the optical
response of the ensemble is broadened by the inhomogeneous
distribution of particle sizes and shapes.  The majority of
particles are off resonance with the exciting laser, and thus have
nonlinear responses much smaller than or even opposite in sign to
the resonant particles.  This leads to an overall measured effect
that is greatly reduced and whose dynamics are obscured.  By
contrast, isolating single particles allows the quantitative
measurement of inherent properties.\cite{Itoh01,Liau01,vanDijk05}

We report the first measurements of resonant nonlinearities of
surface plasmons in single metal nanoparticles, specifically Au
nanorods.\cite{SPIE,arXiv}  We measure a nonlinear scattering
cross-section that is much larger than that obtained from ensembles
of nanorods. Surprisingly, the measured effect can be explained
entirely as the result of heating of conduction electrons, with no
measurably nonlinearity directly associated with coherent plasmon
oscillation.  This indicates that the strong optical driving fields
induce a novel damping and saturation of the plasmonic response.

\section{Experimental}

\begin{figure*}[t]
 \centering
\begin{center}
\includegraphics[width=5.5in]{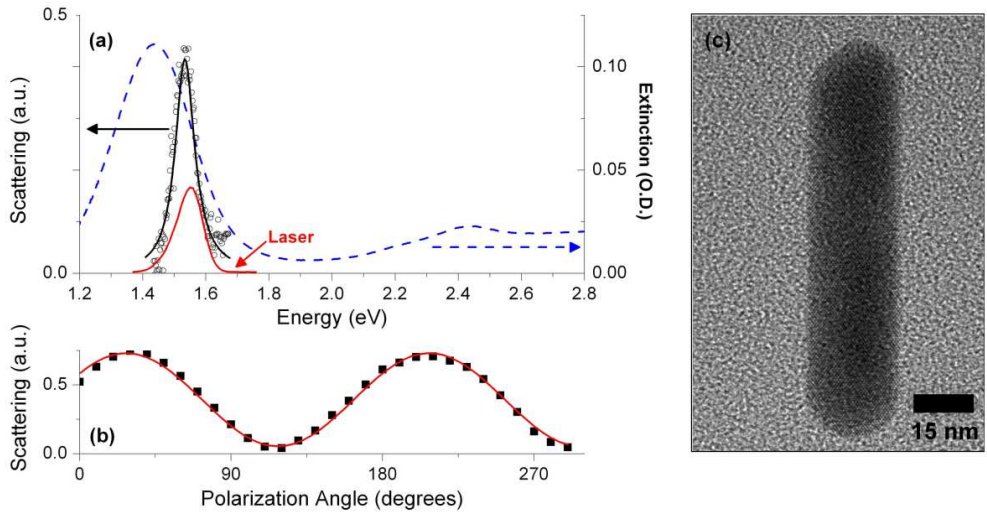}
\caption{\label{spectra} (Color online) (a) Extinction of an
ensemble of Au nanorods in aqueous solution (dashed blue line);
scattering spectrum from a single nanorod on a glass surface
(circles); calculated scattering spectrum for a single rod (solid
black line); and measured spectrum of the laser used to excite the
rods (solid red line). (b) Intensity of laser light scattered off a
single rod as the incident laser polarization is varied (squares),
and sinusoidal fit (red line).  (c) Au nanorod, on a carbon grid,
imaged with a transmission-electron microscope (TEM). }
\end{center}
\end{figure*}

The Au nanorods we study are chemically synthesized using a
seed-mediated growth method.\cite{Nikoobakht03,Liu04} Under the
proper growth conditions, this process produces single-crystal rods
with smooth surfaces, controllable aspect ratios, and $> 95\%$
yield.\cite{Liu05} A transmission-electron-microscope image of a
typical nanorod is shown in Fig. \ref{spectra}(c).  The rods exhibit
a strong longitudinal plasmon resonance, whose frequency is
determined by the aspect ratio of the rods.\cite{Bohren83}  Damping
due to interband transitions is reduced by selecting particles with
a plasmon resonance near 1.55 eV,\cite{Sonnichsen02} which matches
the Ti:Sapphire laser light used to excite and probe the rods.

The gold-nanorod solution is prepared as follows.  Spherical seed
particles are produced by mixing 0.25 mL of 10 mM HAuCl$_4$ solution
with 10 mM of 0.1 M CTAB (cetyltrimethylammoniumbromide) solution at
room temperature, and then quickly injecting 0.6 mL of freshly
prepared 10 mM NaHBH$_4$ solution under vigorous stirring. In order
to grow the seeds into rods, 50 mL of 0.1 M CTAB solution is
prepared and maintained at 28 $^\circ$C. To this solution, 2.5 mL of
10 mM HAuCl$_4$, and 0.5 mL of 10 mM AgNO$_3$ are added.  1.0 mL of
1.0 M HCl is also added, to maintain the stability of the final
produce. Au(III) is reduced to Au(I) by injecting 0.4 mL of 0.1 M
ascorbic acid.  Finally, 0.12 mL of the gold-seed solution is added
to begin the nanorod growth, and growth is completed overnight under
steady stirring. No further size selection is performed.

The sample consists of sparsely dispersed and isolated rods, bound
to a glass coverslip.  The coverslip is first cleaned for 10 minutes
with an equal mixture of 30\% H$_2$O$_2$ and 98\% H$_2$SO$_4$, and
then coated with an MPTMS (3-mercaptopropyltrimethylsilane)
monolayer, using a two-step gas-phase silinazation
procedure.\cite{Jung04}  The functionalized glass substrate is
dipped into the Au-nanorod solution for 30 minutes.  The sample is
then washed thoroughly in deionized water and dried in air.
Atomic-force microscopy (AFM) is used to verify that the rods are
well isolated on the surface, so that individual rods can
subsequently be probed optically.

Optical measurements on the single rods are made using
total-internal-reflection microscopy.\cite{Liu04,Liu05,Sonnichsen00}
Incident light is focussed onto the sample through a glass prism.
Scattered light is collected using a microscope objective and is
imaged onto a multimode optical fiber, which selects a 1.5 $\mu$m
spot on the sample for observation. For spectral measurements, the
light is sent to a spectrometer equipped with a cooled CCD array
detector (Andor); for time-resolved measurements, the light is sent
to an avalanche photodiode (Hamamatsu).

\section{Identification of Single Rods}

Making single-rod measurements requires carefully ensuring that only
one rod at a time is being probed.  The primary method of
identifying single rods is to excite with incoherent white light and
measure the scattering spectrum; a typical spectrum is shown in Fig.
\ref{spectra}(a).  The narrow resonance, much less broad than the
ensemble peak, indicates that the scattering comes from a single
rod.  If two or more nanorods are probed, they will have different
shapes, and thus different plasmon resonance frequencies; the
measured scattering spectrum will then be broader than the
single-rod spectrum.

The scattering spectrum is quantitatively compared to a calculation
in the quasi-static approximation,\cite{Liu04,Liu05,Bohren83} as
shown in Fig. \ref{spectra}(a). We treat the rod as a prolate
ellipsoid, and approximate the asymmetric environment of the rod as
a homogeneous, transparent medium with dielectric constant
$\epsilon_m = 1.3$.  For incident light polarized parallel to the
long axis of the rod, the polarizability of the particle is
\begin{equation}
\label{susceptibility}
\alpha = V \frac{\epsilon -
\epsilon_m}{\epsilon_m+L(\epsilon - \epsilon_m)} \, ,
\end{equation}
where $V$ is the volume of the rod, $\epsilon$ is the dielectric
function of Au, and L is a geometric factor:
\begin{equation}
L = \frac{1-e^2}{e^2} \left( \frac{1}{2e} \ln \frac{1+e}{1-e} -1
\right) \, ,
\end{equation}
where $e$ is the eccentricity of the ellipsoid.  The scattering
cross-section is then given by
\begin{equation}
\label{cross-section}
\sigma_{scat} = \frac{k^4}{6 \pi} \left|
\alpha \right| ^2 \, ,
\end{equation}
where $k$ is the wavenumber of the incident light.

The imaginary part of the dielectric function of Au is taken to
be\cite{Rosei73,Rosei74,Guerrisi75}
\begin{equation}
\epsilon_2 (\omega,T_e) = \frac{\omega_p^2 \gamma(T_e)}{\omega
\left[ \omega_p^2 + \gamma(T_e)^2 \right]} +
\epsilon_2^{d-c}(\omega,T_e) \, , \label{dielectric}
\end{equation}
where $\omega$ is the optical frequency and $T_e$ is the temperature
of the conduction electrons in the rod. The first term is the Drude
free-electron contribution; $\omega_p$ is the bulk plasmon
frequency, and $\gamma$ is the plasmon damping rate. The second term
is the contribution of transitions between the $d$ and the
conduction bands. The real part of the dielectric function is
calculated from this imaginary part using the Kramers-Kronig
relation.  The matrix elements of the interband transitions and the
Drude plasmon frequency are adjusted to reproduce experimental
dielectric functions.\cite{Johnson72}

\begin{figure}[t]
 \centering
\begin{center}
\includegraphics[width=3.375in]{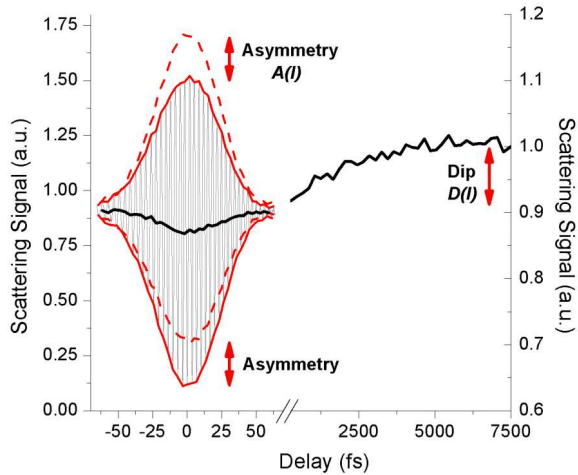}
\caption{\label{interferograms} (Color online) Single-rod scattering
signal as a function of delay between two incident laser pulses.
Left-hand side: scattering intensity for overlapping pulses with an
energy of 47 pJ (light line); envelope of the interference pattern
(solid red line); the same envelope, inverted about the average
scattering signal at a delay of 75 fs (dashed red line); and the
average of the upper and lower envelopes (heavy line). Right-hand
side: scattering intensity for non-overlapping pulses with an energy
of 94 pJ.  Note that both vertical and horizontal scales are
different on the two sides of the graph.}
\end{center}
\end{figure}

In comparing the calculated and measured scattering spectra, the
only free parameter is the aspect ratio of the rod.  For the
particular rod in Fig. \ref{spectra}(a), the fitted aspect ratio is
5.25, consistent with the rod shapes measured by TEM.  A different
choice of refractive index for the surrounding medium changes the
fitted nanorod aspect ratio, but has no other appreciable effect on
calculated optical properties. The very good agreement between the
calculated and measured scattering linewidths is thus a clear
indication that only a single rod is being probed.

Identification of single rods is further supported by the strong
polarization dependence of the scattering, as shown in Fig.
\ref{spectra}(b).  The nearly complete modulation is consistent with
scattering from a single, oriented dipole, rather than multiple
particles.

\section{Ultrafast Nonlinearities}

Nonlinearities of the single nanorods are measured using an
interferometric scattering technique. The rods are excited with
20-fs pulses from a mode-locked, cavity-dumped Ti:Sapphire
laser.\cite{Liau99} The pulses are split into two equal-intensity
parts,\cite{Liau01,Petek98} and the delay of one of the pulses is
controlled relative to the other by moving a retroreflector, using
either a calibrated stepper motor or a piezoelectric transducer.  A
single lens focuses the two pulses to a common 20-$\mu$m spot on the
sample. The signal is processed by a lock-in amplifier, which is
synchronized to a chopper that modulates both laser beams.

\begin{figure}[t]
 \centering
\begin{center}
\includegraphics[width=3.45in]{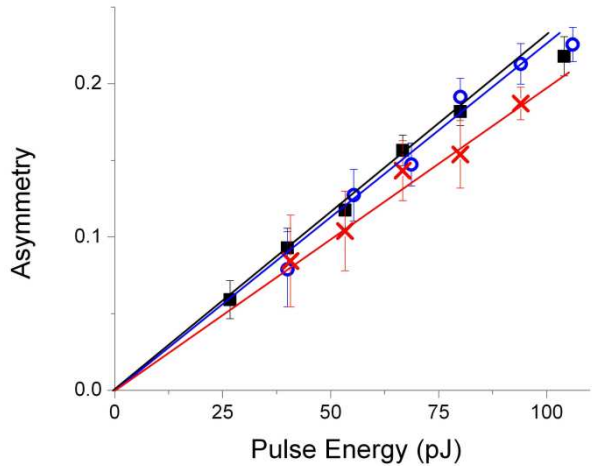}
\caption{\label{pump-power} (Color online) Measured asymmetry in
interference patterns for three different rods.}
\end{center}
\end{figure}

Fig. \ref{interferograms} shows an example of the scattering signal
from a single rod for short delays; the measured interference
pattern exhibits a pronounced asymmetry in intensity. When the laser
pulses interfere constructively, the incident intensity is doubled,
but the amount of scattering from the rod increases by less than a
factor of two, meaning that the scattering cross-section of the rod
is smaller for the higher intensity. The interference patterns do
not change as the repetition rate of the laser is varied, indicating
that slow, cumulative effects are not important.  The nonlinearity
thus arises within the 20-fs pulse duration.

\begin{figure}[t]
 \centering
\begin{center}
\includegraphics[width=3.25in]{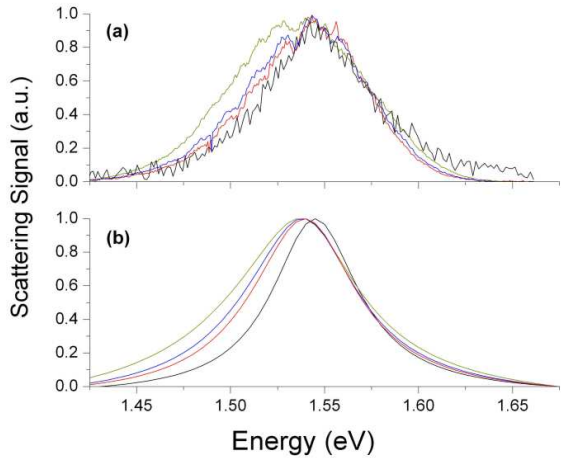}
\caption{\label{nonlinear-spectra} (Color online) (a) Measured
single-rod scattering spectra for different incident intensities.
The rightmost curve is the linear spectrum measured using
incoherent, broadband excitation. The other curves are measured
using a single laser beam, and correspond to pulse energies of 52,
82, and 192 pJ, from right to left.  The spectra are normalized for
ease of comparison.  (b) Corresponding calculated scattering
spectra, assuming spectral changes are due to instantaneous heating
of conduction electrons.}
\end{center}
\end{figure}

These measurements make it possible to establish the magnitude of
the ultrafast nonlinearity.  Figure \ref{pump-power} shows for
different rods the measured asymmetry as a function of the laser
intensity, $I$, in each pulse. The nearly linear dependence
indicates a third-order nonlinearity.  That is, the scattering
cross-section can be written $\sigma (I) = \sigma^{(0)} + I
\sigma^{(3)}$, so that the measured asymmetry should be
\begin{equation}
A(I) = 4 I \frac{\sigma^{(3)}}{\sigma^{(0)}-2 I \sigma^{(3)}} \, .
\end{equation}
As shown in Fig. \ref{pump-power}, this formula fits our data well.

The fit to the data gives a normalized nonlinear cross-section
$\sigma^{(3)}/\sigma^{(0)} = (7.5 \pm 0.9) \times 10^{-11}$
cm$^2$/W. Transient-transmission measurements on nanorod ensembles
in solution show a corresponding average nonlinear cross-section of
approximately $2.5 \times 10^{-13}$ cm$^2$/W;\cite{Park05}  the
significantly lower value is the result of the large number of
non-resonant nanorods, whose nonlinearities are smaller than and can
even oppose those of the resonant rods. The single-rod nonlinear
cross-section also implies a nonlinear susceptibility for Au over
the laser bandwidth of $\chi^{(3)} \approx 5 \times 10^{-18}$
m$^2$/V$^2$.  As a result of this large nonlinearity, the change in
scattering cross-section can reach over 20\% at high laser
intensities. If the laser power is increased further, optical damage
occurs, and the scattering signal gradually and irreversibly
decreases.

Further insight into the measured nonlinearity is obtained by
measuring the dependence of the nanorod scattering spectrum on
incident laser power.  Results for a particular nanorod are shown in
Fig. \ref{nonlinear-spectra}(a); an intensity-dependent red shift
$\Delta \omega$ and line broadening $\Delta \gamma$ are clearly
seen. Both effects are linear in $I$; for this rod, $\Delta \omega =
59$ meV/nJ, and $\Delta \gamma = 87$ meV/nJ.

\section{Picosecond Nonlinearities}

\begin{figure}[t]
 \centering
\begin{center}
\includegraphics[width=3.45in]{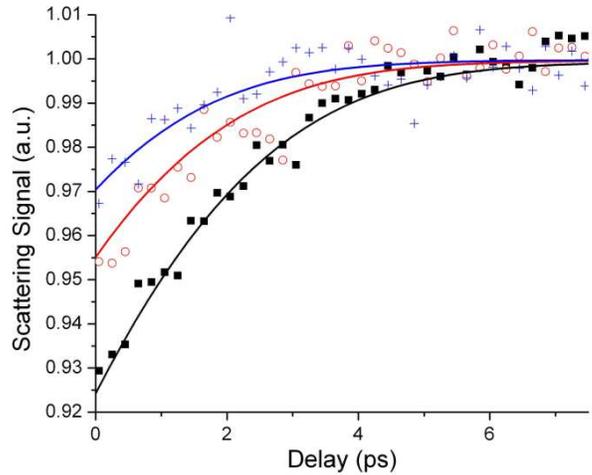}
\caption{\label{electron-phonon} (Color online) Measured single-rod
scattering signal as a function of time delay between two incident
laser pulses, normalized by the measured signal at a delay of 20 ps
(points), and calculated change in scattering (lines). The three
curves, from top to bottom, correspond to pulse energies of 26, 53,
and 94 pJ.}
\end{center}
\end{figure}

Having established the magnitude of the nonlinearity, we next
investigate its time dependence. To do so, we perform measurements
with longer time delays, so that the two laser pulses no longer
overlap. Figure \ref{interferograms} shows a representative result,
and Fig. \ref{electron-phonon} gives similar results for different
laser powers. The response is characteristic of the heating of
conduction electrons by the laser pulse, followed by their cooling
and equilibration with lattice
phonons.\cite{Hodak00,Voisin01,Link03,Park05}  Increasing the delay
up to 150 ps results in no detectable change in the scattering
signal, indicating that effects related to the heating of lattice
phonons are unimportant on experimental time scales. The data in
Fig. \ref{electron-phonon} have been normalized to the signal at
long delays; any ultrafast nonlinearity arising within the laser
pulse duration will result in a change in this reference level, but
will not otherwise not affect the time-delay-dependent data.

The picosecond-scale results can be modeled as follows.  The amount
of light transferred from the first laser pulse to the conduction
electrons is determined by calculating the nanorod absorption
cross-section; this energy transfer results in a higher electron
temperature, $T_e$. The subsequent evolution of $T_e$ is calculated
by treating the conduction electrons and the lattice phonons as two
coupled thermal reservoirs.\cite{Arbouet03,Park05}  The time
evolution of the reservoir temperatures is given by
\begin{eqnarray}
C_e \,T_e'(t) & = &g \left[ T_l(t) - T_e(t) \right] \\
C_l \, T_l'(t) & = & g \left[ T_e(t) - T_l(t) \right] \, ,
\end{eqnarray}
where $C_e$ and $C_l$ are the heat capacities of the electrons and
the lattice, respectively; $T_l$ is the lattice temperature; and $g$
is the electron-phonon coupling coefficient.

The effect of elevated $T_e$ is to alter the dielectric function for
Au, as described by Eqn.
(\ref{dielectric}).\cite{Rosei73,Rosei74,Guerrisi75} From the
dielectric function, the scattering spectrum can be calculated
according to Eqns. (\ref{susceptibility}) and (\ref{cross-section}).
For higher electron temperatures, the plasmon is broadened and
red-shifted. The modified plasmon response is used to calculate the
amount of light scattered from the second laser pulse.

This calculated scattering signal is fit to the data using a single
free parameter, relating the measured laser power to the optical
intensity incident on the rod. Fig. \ref{electron-phonon} shows, for
a single rod, results for three different laser powers.  Equally
good agreement was obtained for several other laser powers and for
other rods, indicating that electron heating can account for the
measured nonlinearity on picosecond time scales.

\section{Plasmon Saturation}

Unexpectedly, the same thermal model that explains the picosecond
measurements also quantitatively explains the measured
nonlinearities on femtosecond time scales. More precisely, we can
extrapolate the measured thermal nonlinearity for a given laser
intensity, $I$, to zero time delay; this gives a ``dip'' $D(I)$ in
the normalized scattering signal. (See Fig. \ref{interferograms}; in
this case $D \approx 9\%$.) The measured values of $D(I)$ can be
compared to the asymmetries, $A(I)$, of the measured interference
patterns. Assuming that the only nonlinearity, even for the shortest
time delays, is due to electron heating, we obtain $A(I) = 2D(2I)$.
(We note that this relation takes into account the dependence of the
reference level for $D(I)$ on the intensity of the two laser
pulses.) We observe exactly this relation, within our experimental
error, meaning that we see no change in the dynamics of the
nonlinearity as we move from femtosecond to picosecond time scales.

This observation is consistent with the increase in the plasmon
linewidth at high pulse energies, as shown in Fig.
\ref{pump-power}(a).  Fig. \ref{pump-power}(b) shows theoretical
scattering spectra for the same pump powers, assuming that the only
spectral changes are due to instantaneous heating of the conduction
electrons.  Differences between the calculations and the
measurements, particularly in the line shapes, are likely due to
femtosecond-scale dynamics that are not captured by the extreme
assumption of instantaneous electron heating. Nonetheless, the good
agreement provides another strong indication that a nearly thermal
distribution of electrons is produced in the rod within a time short
compared to the 20-fs laser pulse duration.

This is unexpected, since the resonant laser pulses should excite
plasmons that remain coherent for 15 fs, based on the measured
linear-scattering linewidth (see Fig. \ref{spectra}(a)). Any initial
nonlinearity would then be due to deviation of the plasmon motion
from perfect harmonic oscillation, caused, for example, by
confinement of electrons by the boundaries of the
rod.\cite{Lippitz05} The amplitude of electron oscillation can be
estimated by considering the dipole moment $D$ induced by the
applied field:
\begin{equation}
D = \alpha E = n d e\, ,
\end{equation}
where $E$ is the applied field, $n$ is the number of electrons in
the rod, $d$ is their displacement, and $e$ is the electronic
charge. Using known material parameters for Au then gives, for pulse
energies of 100 pJ, an electron displacement $d$ approximately 8\%
of the rod length.  The deviation from simple harmonic oscillation
should then be considerable, implying a significant coherent
nonlinearity. Such a mechanism would also be responsible for the
generation of third-harmonic radiation by resonantly-driven
plasmons.\cite{Lamprecht99,Lippitz05}

The absence of any measurable coherent nonlinearity, and the
immediate emergence of an incoherent thermal nonlinearity, indicate
that the plasmon cannot be coherently oscillating over the duration
of the laser pulse.  In other words, the strong, resonant laser
excitation must be responsible for increasing the plasmon damping
rate and destroying its coherence.  The reduction in the plasmon
lifetime then means that it is impossible to resolve any coherent
nonlinearities with the 20-fs pulses used.

The increased damping may be due to a greater rate of dephasing
collisions with the nanorod boundaries or to higher-order
plasmon-plasmon or plasmon-electron interactions. Such interactions
are also manifest in the significant electron energies that are
observed in photoemission measurements when plasmons are resonantly
excited.\cite{Banfi03,Kubo05}
\\

\section{Conclusions}

The measurements described in this paper have established for the
first time the magnitude of resonant optical nonlinearities in
single Au nanorods. A benchmark value of 20\% has been obtained for
the nonlinear change in the scattering cross-section, using pulse
energies that induce no optical damage. Surprisingly, resonant
excitation of plasmons results in the same nonlinearity as
incoherent excitation of conduction electrons. This indicates that
strongly driven plasmons experience a new, intensity-dependent
damping. There are still several potential routes towards achieving
stronger nonlinearities in these systems, such as embedding the
nanorods in a polarizable medium, or assembling them into ordered
structures. Our current observations thus represent a first step
towards achieving very large optical nonlinearities on the nanometer
scale.

\section*{Acknowledgements}

We thank Dr. A. Bakhtyari for valuable assistance and Prof. H. Petek
for helpful discussions. This work was principally supported by the
MRSEC program of the NSF under grant No. DMR 0213745, with
additional support from the UC-ANL Consortium for Nanoscience
Research and from NSF grant No. CHE 0317009. M.P. is supported by
the Grainger Postdoctoral Fellowship in Experimental Physics from
the University of Chicago.

\bibliography{rods}

\begin{thebibliography}{37}
\expandafter\ifx\csname natexlab\endcsname\relax\def\natexlab#1{#1}\fi
\expandafter\ifx\csname bibnamefont\endcsname\relax
  \def\bibnamefont#1{#1}\fi
\expandafter\ifx\csname bibfnamefont\endcsname\relax
  \def\bibfnamefont#1{#1}\fi
\expandafter\ifx\csname citenamefont\endcsname\relax
  \def\citenamefont#1{#1}\fi
\expandafter\ifx\csname url\endcsname\relax
  \def\url#1{\texttt{#1}}\fi
\expandafter\ifx\csname urlprefix\endcsname\relax\def\urlprefix{URL }\fi
\providecommand{\bibinfo}[2]{#2}
\providecommand{\eprint}[2][]{\url{#2}}

\bibitem[{\citenamefont{Saleh and Teich}(1991)}]{Saleh91}
\bibinfo{author}{\bibfnamefont{B.~E.~A.} \bibnamefont{Saleh}} \bibnamefont{and}
  \bibinfo{author}{\bibfnamefont{M.~C.} \bibnamefont{Teich}},
  \emph{\bibinfo{title}{{Fundamentals of Photonics}}} (\bibinfo{publisher}{John
  Wiley \& Sons}, \bibinfo{address}{New York}, \bibinfo{year}{1991}).

\bibitem[{\citenamefont{Barnes et~al.}(2003)\citenamefont{Barnes, Dereux, and
  Ebbesen}}]{Barnes03}
\bibinfo{author}{\bibfnamefont{W.~L.} \bibnamefont{Barnes}},
  \bibinfo{author}{\bibfnamefont{A.}~\bibnamefont{Dereux}}, \bibnamefont{and}
  \bibinfo{author}{\bibfnamefont{T.~W.} \bibnamefont{Ebbesen}},
  \bibinfo{journal}{Nature} \textbf{\bibinfo{volume}{424}},
  \bibinfo{pages}{824} (\bibinfo{year}{2003}).

\bibitem[{\citenamefont{Maier and Atwater}(2005)}]{Maier05}
\bibinfo{author}{\bibfnamefont{S.~A.} \bibnamefont{Maier}} \bibnamefont{and}
  \bibinfo{author}{\bibfnamefont{H.~A.} \bibnamefont{Atwater}},
  \bibinfo{journal}{J. Appl. Phys.} \textbf{\bibinfo{volume}{98}},
  \bibinfo{pages}{011101} (\bibinfo{year}{2005}).

\bibitem[{\citenamefont{Quinten et~al.}(1998)\citenamefont{Quinten, Leitner,
  Krenn, and Aussenegg}}]{Quinten98}
\bibinfo{author}{\bibfnamefont{M.}~\bibnamefont{Quinten}},
  \bibinfo{author}{\bibfnamefont{A.}~\bibnamefont{Leitner}},
  \bibinfo{author}{\bibfnamefont{J.~R.} \bibnamefont{Krenn}}, \bibnamefont{and}
  \bibinfo{author}{\bibfnamefont{F.~R.} \bibnamefont{Aussenegg}},
  \bibinfo{journal}{Opt. Lett.} \textbf{\bibinfo{volume}{23}},
  \bibinfo{pages}{1331} (\bibinfo{year}{1998}).

\bibitem[{\citenamefont{{S. A. Maier {\it et al.}}}(2003)}]{Maier03}
\bibinfo{author}{\bibnamefont{{S. A. Maier {\it et al.}}}},
  \bibinfo{journal}{Nature Materials} \textbf{\bibinfo{volume}{2}},
  \bibinfo{pages}{229} (\bibinfo{year}{2003}).

\bibitem[{\citenamefont{Nie and Emory}(1997)}]{Nie97}
\bibinfo{author}{\bibfnamefont{S.}~\bibnamefont{Nie}} \bibnamefont{and}
  \bibinfo{author}{\bibfnamefont{S.~R.} \bibnamefont{Emory}},
  \bibinfo{journal}{Science} \textbf{\bibinfo{volume}{275}},
  \bibinfo{pages}{1102} (\bibinfo{year}{1997}).

\bibitem[{\citenamefont{{K. Kneipp {\it et al.}}}(1997)}]{Kneipp97}
\bibinfo{author}{\bibnamefont{{K. Kneipp {\it et al.}}}},
  \bibinfo{journal}{Phys. Rev. Lett.} \textbf{\bibinfo{volume}{78}},
  \bibinfo{pages}{1667} (\bibinfo{year}{1997}).

\bibitem[{\citenamefont{{M. J. Feldstein {\it et al.}}}(1997)}]{Feldstein97}
\bibinfo{author}{\bibnamefont{{M. J. Feldstein {\it et al.}}}},
  \bibinfo{journal}{J. Am. Chem. Soc.} \textbf{\bibinfo{volume}{119}},
  \bibinfo{pages}{6638} (\bibinfo{year}{1997}).

\bibitem[{\citenamefont{Hodak et~al.}(2000)\citenamefont{Hodak, Henglein, and
  Hartland}}]{Hodak00}
\bibinfo{author}{\bibfnamefont{J.~H.} \bibnamefont{Hodak}},
  \bibinfo{author}{\bibfnamefont{A.}~\bibnamefont{Henglein}}, \bibnamefont{and}
  \bibinfo{author}{\bibfnamefont{G.~V.} \bibnamefont{Hartland}},
  \bibinfo{journal}{J. Phys. Chem. B} \textbf{\bibinfo{volume}{104}},
  \bibinfo{pages}{9954} (\bibinfo{year}{2000}).

\bibitem[{\citenamefont{Voisin et~al.}(2001)\citenamefont{Voisin, Fatti,
  Christofilos, and Vall{\'e}e}}]{Voisin01}
\bibinfo{author}{\bibfnamefont{C.}~\bibnamefont{Voisin}},
  \bibinfo{author}{\bibfnamefont{N.~D.} \bibnamefont{Fatti}},
  \bibinfo{author}{\bibfnamefont{D.}~\bibnamefont{Christofilos}},
  \bibnamefont{and}
  \bibinfo{author}{\bibfnamefont{F.}~\bibnamefont{Vall{\'e}e}},
  \bibinfo{journal}{J. Phys. Chem. B} \textbf{\bibinfo{volume}{105}},
  \bibinfo{pages}{2264} (\bibinfo{year}{2001}).

\bibitem[{\citenamefont{Link and El-Sayed}(2003)}]{Link03}
\bibinfo{author}{\bibfnamefont{S.}~\bibnamefont{Link}} \bibnamefont{and}
  \bibinfo{author}{\bibfnamefont{M.}~\bibnamefont{El-Sayed}},
  \bibinfo{journal}{Annu. Rev. Phys. Chem.} \textbf{\bibinfo{volume}{54}},
  \bibinfo{pages}{331} (\bibinfo{year}{2003}).

\bibitem[{\citenamefont{{A. Arbouet {\it et al.}}}(2003)}]{Arbouet03}
\bibinfo{author}{\bibnamefont{{A. Arbouet {\it et al.}}}},
  \bibinfo{journal}{Phys. Rev. Lett.} \textbf{\bibinfo{volume}{90}},
  \bibinfo{pages}{177401} (\bibinfo{year}{2003}).

\bibitem[{\citenamefont{{S. Link {\it et al.}}}(2000)}]{Link00}
\bibinfo{author}{\bibnamefont{{S. Link {\it et al.}}}}, \bibinfo{journal}{Phys.
  Rev. B} \textbf{\bibinfo{volume}{61}}, \bibinfo{pages}{6086}
  (\bibinfo{year}{2000}).

\bibitem[{\citenamefont{{M. Hu {\it et al.}}}(2003)}]{Hu03}
\bibinfo{author}{\bibnamefont{{M. Hu {\it et al.}}}}, \bibinfo{journal}{J. Am.
  Chem. Soc.} \textbf{\bibinfo{volume}{125}}, \bibinfo{pages}{14925}
  (\bibinfo{year}{2003}).

\bibitem[{\citenamefont{Itoh et~al.}(2001)\citenamefont{Itoh, Asahi, and
  Masuhara}}]{Itoh01}
\bibinfo{author}{\bibfnamefont{T.}~\bibnamefont{Itoh}},
  \bibinfo{author}{\bibfnamefont{T.}~\bibnamefont{Asahi}}, \bibnamefont{and}
  \bibinfo{author}{\bibfnamefont{H.}~\bibnamefont{Masuhara}},
  \bibinfo{journal}{Appl. Phys. Lett.} \textbf{\bibinfo{volume}{79}},
  \bibinfo{pages}{1667} (\bibinfo{year}{2001}).

\bibitem[{\citenamefont{Liau et~al.}(2001)\citenamefont{Liau, Unterreiner,
  Chang, and Scherer}}]{Liau01}
\bibinfo{author}{\bibfnamefont{Y.-H.} \bibnamefont{Liau}},
  \bibinfo{author}{\bibfnamefont{A.~N.} \bibnamefont{Unterreiner}},
  \bibinfo{author}{\bibfnamefont{Q.}~\bibnamefont{Chang}}, \bibnamefont{and}
  \bibinfo{author}{\bibfnamefont{N.~F.} \bibnamefont{Scherer}},
  \bibinfo{journal}{J. Phys. Chem. B} \textbf{\bibinfo{volume}{105}},
  \bibinfo{pages}{2135} (\bibinfo{year}{2001}).

\bibitem[{\citenamefont{van Dijk et~al.}(2005)\citenamefont{van Dijk, Lippitz,
  and Orrit}}]{vanDijk05}
\bibinfo{author}{\bibfnamefont{M.~A.} \bibnamefont{van Dijk}},
  \bibinfo{author}{\bibfnamefont{M.}~\bibnamefont{Lippitz}}, \bibnamefont{and}
  \bibinfo{author}{\bibfnamefont{M.}~\bibnamefont{Orrit}},
  \bibinfo{journal}{Phys. Rev. Lett.} \textbf{\bibinfo{volume}{95}},
  \bibinfo{pages}{267406} (\bibinfo{year}{2005}).

\bibitem[{\citenamefont{Pelton et~al.}(2005)\citenamefont{Pelton, Liu, Park,
  Scherer, and Guyot-Sionnest}}]{SPIE}
\bibinfo{author}{\bibfnamefont{M.}~\bibnamefont{Pelton}},
  \bibinfo{author}{\bibfnamefont{M.}~\bibnamefont{Liu}},
  \bibinfo{author}{\bibfnamefont{S.}~\bibnamefont{Park}},
  \bibinfo{author}{\bibfnamefont{N.~F.} \bibnamefont{Scherer}},
  \bibnamefont{and}
  \bibinfo{author}{\bibfnamefont{P.}~\bibnamefont{Guyot-Sionnest}},
  \bibinfo{journal}{Proc. SPIE} \textbf{\bibinfo{volume}{5927}},
  \bibinfo{pages}{191} (\bibinfo{year}{2005}).

\bibitem[{\citenamefont{Pelton et~al.}()\citenamefont{Pelton, Liu, Park,
  Scherer, and Guyot-Sionnest}}]{arXiv}
\bibinfo{author}{\bibfnamefont{M.}~\bibnamefont{Pelton}},
  \bibinfo{author}{\bibfnamefont{M.}~\bibnamefont{Liu}},
  \bibinfo{author}{\bibfnamefont{S.}~\bibnamefont{Park}},
  \bibinfo{author}{\bibfnamefont{N.~F.} \bibnamefont{Scherer}},
  \bibnamefont{and}
  \bibinfo{author}{\bibfnamefont{P.}~\bibnamefont{Guyot-Sionnest}},
  \bibinfo{note}{http://arxiv.org/abs/cond-mat/0506158}.

\bibitem[{\citenamefont{Nikoobakht and El-Sayed}(2003)}]{Nikoobakht03}
\bibinfo{author}{\bibfnamefont{B.}~\bibnamefont{Nikoobakht}} \bibnamefont{and}
  \bibinfo{author}{\bibfnamefont{M.~A.} \bibnamefont{El-Sayed}},
  \bibinfo{journal}{Chem. Mater.} \textbf{\bibinfo{volume}{15}},
  \bibinfo{pages}{1957} (\bibinfo{year}{2003}).

\bibitem[{\citenamefont{Liu and Guyot-Sionnest}(2004)}]{Liu04}
\bibinfo{author}{\bibfnamefont{M.}~\bibnamefont{Liu}} \bibnamefont{and}
  \bibinfo{author}{\bibfnamefont{P.}~\bibnamefont{Guyot-Sionnest}},
  \bibinfo{journal}{J. Phys. Chem. B} \textbf{\bibinfo{volume}{108}},
  \bibinfo{pages}{5882} (\bibinfo{year}{2004}).

\bibitem[{\citenamefont{Liu and Guyot-Sionnest}(2005)}]{Liu05}
\bibinfo{author}{\bibfnamefont{M.}~\bibnamefont{Liu}} \bibnamefont{and}
  \bibinfo{author}{\bibfnamefont{P.}~\bibnamefont{Guyot-Sionnest}},
  \bibinfo{journal}{J. Phys. Chem. B} \textbf{\bibinfo{volume}{109}},
  \bibinfo{pages}{22192} (\bibinfo{year}{2005}).

\bibitem[{\citenamefont{Bohren and Huffman}(1983)}]{Bohren83}
\bibinfo{author}{\bibfnamefont{C.~F.} \bibnamefont{Bohren}} \bibnamefont{and}
  \bibinfo{author}{\bibfnamefont{D.~R.} \bibnamefont{Huffman}},
  \emph{\bibinfo{title}{{Absorption and Scattering of Light by Small
  Particles}}} (\bibinfo{publisher}{John Wiley \& Sons}, \bibinfo{address}{New
  York}, \bibinfo{year}{1983}).

\bibitem[{\citenamefont{{C. S{\"o}nnichsen {\it et al.}}}(2002)}]{Sonnichsen02}
\bibinfo{author}{\bibnamefont{{C. S{\"o}nnichsen {\it et al.}}}},
  \bibinfo{journal}{Phys. Rev. Lett.} \textbf{\bibinfo{volume}{88}},
  \bibinfo{pages}{077402} (\bibinfo{year}{2002}).

\bibitem[{\citenamefont{{H. Jung {\it at al.}}}(2004)}]{Jung04}
\bibinfo{author}{\bibnamefont{{H. Jung {\it at al.}}}}, \bibinfo{journal}{Nano
  Lett.} \textbf{\bibinfo{volume}{4}}, \bibinfo{pages}{2171}
  (\bibinfo{year}{2004}).

\bibitem[{\citenamefont{{C. S{\"o}nnichsen {\it el al.}}}(2000)}]{Sonnichsen00}
\bibinfo{author}{\bibnamefont{{C. S{\"o}nnichsen {\it el al.}}}},
  \bibinfo{journal}{Appl. Phys. Lett.} \textbf{\bibinfo{volume}{77}},
  \bibinfo{pages}{2949} (\bibinfo{year}{2000}).

\bibitem[{\citenamefont{Rosei et~al.}(1973)\citenamefont{Rosei, Antonangeli,
  and Grassano}}]{Rosei73}
\bibinfo{author}{\bibfnamefont{R.}~\bibnamefont{Rosei}},
  \bibinfo{author}{\bibfnamefont{F.}~\bibnamefont{Antonangeli}},
  \bibnamefont{and} \bibinfo{author}{\bibfnamefont{U.~M.}
  \bibnamefont{Grassano}}, \bibinfo{journal}{Surf. Sci.}
  \textbf{\bibinfo{volume}{37}}, \bibinfo{pages}{689} (\bibinfo{year}{1973}).

\bibitem[{\citenamefont{Rosei}(1974)}]{Rosei74}
\bibinfo{author}{\bibfnamefont{R.}~\bibnamefont{Rosei}},
  \bibinfo{journal}{Phys. Rev. B} \textbf{\bibinfo{volume}{10}},
  \bibinfo{pages}{474} (\bibinfo{year}{1974}).

\bibitem[{\citenamefont{Guerrisi et~al.}(1975)\citenamefont{Guerrisi, Rosei,
  and Winsemius}}]{Guerrisi75}
\bibinfo{author}{\bibfnamefont{M.}~\bibnamefont{Guerrisi}},
  \bibinfo{author}{\bibfnamefont{R.}~\bibnamefont{Rosei}}, \bibnamefont{and}
  \bibinfo{author}{\bibfnamefont{P.}~\bibnamefont{Winsemius}},
  \bibinfo{journal}{Phys. Rev. B} \textbf{\bibinfo{volume}{12}},
  \bibinfo{pages}{557} (\bibinfo{year}{1975}).

\bibitem[{\citenamefont{Johnson and Christy}(1972)}]{Johnson72}
\bibinfo{author}{\bibfnamefont{P.~B.} \bibnamefont{Johnson}} \bibnamefont{and}
  \bibinfo{author}{\bibfnamefont{R.~W.} \bibnamefont{Christy}},
  \bibinfo{journal}{Phys. Rev. B} \textbf{\bibinfo{volume}{6}},
  \bibinfo{pages}{4370} (\bibinfo{year}{1972}).

\bibitem[{\citenamefont{Liau et~al.}(1999)\citenamefont{Liau, Unterreiner,
  Arnett, and Scherer}}]{Liau99}
\bibinfo{author}{\bibfnamefont{Y.-H.} \bibnamefont{Liau}},
  \bibinfo{author}{\bibfnamefont{A.~N.} \bibnamefont{Unterreiner}},
  \bibinfo{author}{\bibfnamefont{D.~C.} \bibnamefont{Arnett}},
  \bibnamefont{and} \bibinfo{author}{\bibfnamefont{N.~F.}
  \bibnamefont{Scherer}}, \bibinfo{journal}{Appl. Opt.}
  \textbf{\bibinfo{volume}{38}}, \bibinfo{pages}{7386} (\bibinfo{year}{1999}).

\bibitem[{\citenamefont{Petek and Ogawa}(1998)}]{Petek98}
\bibinfo{author}{\bibfnamefont{H.}~\bibnamefont{Petek}} \bibnamefont{and}
  \bibinfo{author}{\bibfnamefont{S.}~\bibnamefont{Ogawa}},
  \bibinfo{journal}{Prog. Surf. Sci.} \textbf{\bibinfo{volume}{56}},
  \bibinfo{pages}{239} (\bibinfo{year}{1998}).

\bibitem[{\citenamefont{Park et~al.}()\citenamefont{Park, Pelton, Liu,
  Guyot-Sionnest, and Scherer}}]{Park05}
\bibinfo{author}{\bibfnamefont{S.}~\bibnamefont{Park}},
  \bibinfo{author}{\bibfnamefont{M.}~\bibnamefont{Pelton}},
  \bibinfo{author}{\bibfnamefont{M.}~\bibnamefont{Liu}},
  \bibinfo{author}{\bibfnamefont{P.}~\bibnamefont{Guyot-Sionnest}},
  \bibnamefont{and} \bibinfo{author}{\bibfnamefont{N.~F.}
  \bibnamefont{Scherer}}, \bibinfo{note}{in preparation}.

\bibitem[{\citenamefont{Lippitz et~al.}(2005)\citenamefont{Lippitz, van Dijk,
  and Orrit}}]{Lippitz05}
\bibinfo{author}{\bibfnamefont{M.}~\bibnamefont{Lippitz}},
  \bibinfo{author}{\bibfnamefont{M.~A.} \bibnamefont{van Dijk}},
  \bibnamefont{and} \bibinfo{author}{\bibfnamefont{M.}~\bibnamefont{Orrit}},
  \bibinfo{journal}{Nano. Lett.} \textbf{\bibinfo{volume}{5}},
  \bibinfo{pages}{799} (\bibinfo{year}{2005}).

\bibitem[{\citenamefont{Lamprecht et~al.}(1999)\citenamefont{Lamprecht, Krenn,
  Leitner, and Aussenegg}}]{Lamprecht99}
\bibinfo{author}{\bibfnamefont{B.}~\bibnamefont{Lamprecht}},
  \bibinfo{author}{\bibfnamefont{J.~R.} \bibnamefont{Krenn}},
  \bibinfo{author}{\bibfnamefont{A.}~\bibnamefont{Leitner}}, \bibnamefont{and}
  \bibinfo{author}{\bibfnamefont{F.~R.} \bibnamefont{Aussenegg}},
  \bibinfo{journal}{Phys. Rev. Lett.} \textbf{\bibinfo{volume}{83}},
  \bibinfo{pages}{4421} (\bibinfo{year}{1999}).

\bibitem[{\citenamefont{Banfi et~al.}(2003)\citenamefont{Banfi, Ferrini, Peloi,
  and Parmigiani}}]{Banfi03}
\bibinfo{author}{\bibfnamefont{G.}~\bibnamefont{Banfi}},
  \bibinfo{author}{\bibfnamefont{G.}~\bibnamefont{Ferrini}},
  \bibinfo{author}{\bibfnamefont{M.}~\bibnamefont{Peloi}}, \bibnamefont{and}
  \bibinfo{author}{\bibfnamefont{F.}~\bibnamefont{Parmigiani}},
  \bibinfo{journal}{Phys. Rev. B} \textbf{\bibinfo{volume}{67}},
  \bibinfo{pages}{035428} (\bibinfo{year}{2003}).

\bibitem[{\citenamefont{Kubo et~al.}(2005)\citenamefont{Kubo, Onda, Petek, Sun,
  Jung, and Kim}}]{Kubo05}
\bibinfo{author}{\bibfnamefont{A.}~\bibnamefont{Kubo}},
  \bibinfo{author}{\bibfnamefont{K.}~\bibnamefont{Onda}},
  \bibinfo{author}{\bibfnamefont{H.}~\bibnamefont{Petek}},
  \bibinfo{author}{\bibfnamefont{Z.}~\bibnamefont{Sun}},
  \bibinfo{author}{\bibfnamefont{Y.~S.} \bibnamefont{Jung}}, \bibnamefont{and}
  \bibinfo{author}{\bibfnamefont{H.~K.} \bibnamefont{Kim}},
  \bibinfo{journal}{Nano Lett.} \textbf{\bibinfo{volume}{5}},
  \bibinfo{pages}{1123} (\bibinfo{year}{2005}).

\end{thebibliography}
\bibliographystyle{apsrev}

\end{document}